\documentclass[letterpaper]{article}
\usepackage{amsmath}
\usepackage{amssymb}
\usepackage{natbib,alifeconf}  
\usepackage{url,hyperref,cleveref}
\usepackage{booktabs}
\usepackage{glossaries}
\usepackage{adjustbox}
\usepackage{xcolor}
\usepackage{enumitem}
\usepackage{epigraph}

\usepackage{tikz-cd}
\usepackage[normalem]{ulem}

\title{ALIFE2024 template}

\title{Closing the Loop: How Semantic Closure Enables Open-Ended Evolution?}

\author{
    Amahury J. López-Díaz 
    \and
    Carlos Gershenson \\
    \mbox{}\\
    School of Systems Science and Industrial Engineering, Binghamton University, 4400 Vestal Pkwy E, Binghamton, NY 13902, USA\\
    alpez@binghamton.edu
} 

\makeglossaries

\newglossaryentry{rb}
{
        name=relational biology,
        description={A theoretical framework that emphasizes the primacy of relationships and interactions over individual entities, pioneered by Robert Rosen, to understand the organization of biological systems as networks of functional relations}
}

\newglossaryentry{pb}
{
        name=physical biosemiotics,
        description={An approach developed by Howard Pattee, which studies the interplay between symbolic information (e.g., genetic code) and physical constraints (e.g., biochemical reactions), emphasizing the fundamental role of symbols in biological systems}
}

\newglossaryentry{ep}
{
        name=ecological psychology,
        description={A branch of psychology focusing on how organisms perceive and act upon their environments, pioneered by J.J. Gibson, emphasizing the direct perception of affordances in the environment}
}

\newglossaryentry{ce}
{
        name=computational enactivism,
        description={A perspective combining principles of enactive cognition with computational models, focusing on how autonomous agents actively generate meaning through their interactions with the environment}
}

\newglossaryentry{MR}
{
        name=(Metabolism{,}~Repair)-systems,
        description={(M, R)-systems are relational models developed by Robert Rosen, where “M” stands for Metabolism and “R” stands for Repair, representing biological systems as sets of functional components organized in mutual causal loops to maintain themselves}
}

\newglossaryentry{FA}
{
        name=(Fabrication{,}~Assembly)-systems,
        description={(F, A)-systems are an extension of Rosen's (Metabolic, Repair)-systems, proposed by Jan-Hendrik Hofmeyr, where “F” stands for Fabrication and “A” for Assembly, emphasizing the construction and self-maintenance capabilities in biological systems}
}

\newglossaryentry{rr}
{
        name=relevance realization,
        description={The fundamental process by which organisms transform ill-defined problems into well-defined ones by determining which environmental cues are pertinent to their survival and functioning. The problem of relevance has also been called the problem of meaning, the symbol grounding problem, and the frame problem}
}

\newglossaryentry{material}
{
        name=material cause,
        description={In metabolic terms, the substances and resources necessary for a biological process or structure, constituting the raw materials involved in biological reactions}
}

\newglossaryentry{efficient}
{
        name=efficient cause,
        description={Refers to the mechanisms or catalysts (e.g., enzymes) that directly facilitate biological processes, enabling reactions and transformations to occur}
}

\newglossaryentry{final}
{
        name=final cause,
        description={The purpose or functional end-state of biological systems, interpreted in metabolic terms as the goal-directed maintenance and survival of the organism}
}

\newglossaryentry{formal}
{
        name=formal cause,
        description={The organizational pattern or structural configuration of a biological system, determining how components interact and processes unfold, thus shaping the identity and function of the organism}
}

\newglossaryentry{affordance}
{
        name=affordance,
        description={An ecological psychology term introduced by J.J. Gibson, referring to the actionable possibilities that an environment offers to an organism, inherently related to the organism's capabilities and goals}
}

\newglossaryentry{impredicativity}
{
        name=impredicativity,
        description={A logical condition where an entity or concept is defined in a self-referential way, typically leading to circular dependencies; important in biological and cognitive systems where processes often define themselves recursively}
}

\newglossaryentry{cc}
{
        name = concrete category,
        description={A concrete category is a pair $(\mathcal{C},U)$ such that $\mathcal{C}$ is a \gls{category} and $U:\mathcal{C}\to\textbf{Set}$ (the category of sets and functions) is a \gls{ff}}
}

\newglossaryentry{category}
{
        name=category,
        description={A category $\mathcal{C}$ consists of (i) A collection of objects, denoted as $\text{Ob}(\mathcal{C})$; (ii) For each pair of objects $A, B\in\text{Ob}(\mathcal{C})$, a collection of arrows (also called morphisms) from $A$ to $B$, denoted as $H(A, B)$. This can be empty, but in general there can be multiple, sometimes infinite, arrows between objects. Thus, an arrow is (in general) not reducible to the pair of objects it connects; (iii) A binary operation $\circ: H(B,  C)\times H(A, B)\to H(A, C)$, called composition of morphisms, such that for any three objects $A$, $B$, and $C$, we have associativity and the existence of an identity morphism for every object of $\mathcal{C}$}
}

\newglossaryentry{ff}
{
        name=faithful functor,
        description={Let $\mathcal{C}$ and $\mathcal{D}$ categories and let $F:\mathcal{C}\to\mathcal{D}$ be a functor from $\mathcal{C}$ to $\mathcal{D}$. The functor induces a function $F_{X, Y}: H(X, Y)\to H(F(X), F(Y))$ for every $X, Y\in \text{Ob}(\mathcal{C})$. The functor $F$ is said to be faithful if $F_{X, Y}$ is injective}
}

\begin{document}

\maketitle


\begin{abstract}
This manuscript explores the evolutionary emergence of semantic closure---the self-referential mechanism through which symbols actively construct and interpret their own functional contexts---by integrating concepts from \gls{rb}, \gls{pb}, and \gls{ep} into a unified \gls{ce} framework. By extending Hofmeyr's \Gls{FA}---a continuation of Rosen's \Gls{MR}---with a temporal parametrization, we develop a model capable of capturing critical properties of life, including autopoiesis, anticipation, and adaptation. Our stepwise model traces the evolution of semantic closure from simple reaction networks that recognize regular languages to self-constructing chemical systems with anticipatory capabilities, identifying self-reference as necessary for robust self-replication and open-ended evolution. Such a computational enactivist perspective underscores the essential necessity of implementing syntax-pragmatic transformations into realizations of life, providing a cohesive theoretical basis for a recently proposed trialectic between autopoiesis, anticipation, and adaptation to solve the problem of \gls{rr}. Thus, our work opens avenues for new models of computation that can better capture the dynamics of life, naturalize agency and cognition, and provide fundamental principles underlying biological information processing.

\end{abstract}

\section{Introduction} 
\epigraph{“How, therefore, we must ask, is it possible for us to distinguish the living from the lifeless if we can describe both conceptually by the motion of inorganic corpuscles?”}{\citealp{pearson1892grammar}}

The question posed by Karl Pearson over a century ago remains central to the philosophical and scientific discourse concerning the nature of life and its differentiation from non-life. Historically, science has attempted to resolve this conundrum through increasingly sophisticated physical and mathematical models. Early attempts, grounded in classical and later quantum physics, explored the possibility of undiscovered laws or unique features intrinsic to living matter \citep{schrodinger1946life, pattee1969physical}. The mid-20th century explosion in molecular biology and origin-of-life research, culminating in the discovery of DNA’s structure and genetic mechanisms, fostered the belief that conventional physical and chemical laws were sufficient to explain life. This view, championed by \citet{kendrew1967phage}, marked the entrenchment of reductionism in theoretical biology. Yet, Pearson's fundamental inquiry persists partially unanswered.

More recently, complex systems science~\citep{Bar-Yam1997, Mitchell:2009, ComplexityExplained, Ladyman2020} has emerged as an alternative paradigm to explain life. Complex systems science seeks universal principles across domains: from spin glasses and sandpiles to cells and societies. According to this narrative, life is an emergent autonomous pattern, produced by the non-linear interactions among its components. However, complex system science does not discriminate life from non-life in physical or functional terms. Furthermore, regardless of the many complex systems science models---nonlinear dynamics~\citep{strogatz2024nonlinear}, ergodic theory~\citep{walters2000introduction}, self-organized criticality~\citep{bak1988self}, agent-based models~\citep{railsback2019agent}, networks~\citep{newman2018networks}, game theory~\citep{fudenberg1991game}, among others---we have developed, rule-based self-organization can only yield limited novelty, which contrasts drastically with the open-ended evolutionary potential of life to innovate using circumscribed resources.

As a complementary viewpoint to complex systems science, physical biosemiotics (pioneered by Howard Pattee) offers a radically different account of what distinguishes living from non-living systems~\citep{pattee2019simulations}. Unlike models that seek universal principles across domains through rule-based self-organization or statistical dynamics, Pattee argued that life originates in a fundamental complementarity between symbolic representation and physical dynamics~\citep{pattee2007necessity}. At the heart of this framework lies the concept of the epistemic cut: a necessary demarcation between quiescent, rate-independent symbolic structures and the rate-dependent, energy-governed dynamics they control~\citep{pattee2001physics}. This distinction, which does not correspond to an ontological dualism but an epistemic necessity, reframes the origin of biological order as dependent not only on physical laws, but on the emergence of symbol-mediated constraints capable of controlling those laws. Pattee's work thus sets the stage for a theory of life in which measurement, control, and memory are not secondary features but primary requirements for evolvability and functional organization.

This leads us to the concept of symbol. Biological systems are pervaded by codes, from the genetic code to behavioral signals \citep{barbieri2003organic, barbieri2018code}. However, symbols do not exist in isolation, but are always part of a semiotic system \citep{pattee1969physical, pattee2001physics}. A semiotic system consist of (i) discrete, rate-independent symbol vehicles; (ii) interpreting structures---non-integrable constraints or codes; and (iii) organizations in which symbols and interpreting structures have functional value \citep{pattee2012does, pattee1986universal}. These semiotic components conform to the pragmatic theory of communication developed by \citet{morris1946signs}, which distinguishes three irreducible dimensions: syntax (formal relations between symbols), semantics (correspondence with referents), and pragmatics (usefulness or action-invoking roles). 

As shown by \citet{cariani1989design}, the three irreducible semiotic axes are essential in understanding how organisms process and act on information. Still, such an irreducibility challenges our conception of modeling, since we are used to reducing everything to symbols and their syntax. In fact, the whole branch of artificial life is based on this premise~\citep{pattee2019simulations}. The idea of reducing everything to syntax is not far-fetched. If we identify observables with symbols and their dynamical laws with syntax, then interpreting structures acquired the role of non-holonomic (or non-integrable) constraints: flexible structures that control dynamics without eliminating configurational alternatives. These constraints, such as mechanical governors in machines or enzymes in cells, cannot be separated from the dynamics they influence. Unlike rigid constraints that merely reduce degrees of freedom, non-integrable constraints retain degeneracy, enabling symbolic control without eliminating the system’s potential variability. They serve as memory readers that bridge the epistemic cut by coupling quiescent symbols with rate-dependent trajectories \citep{pattee1969physical}.

Importantly, constraints are epistemically but not ontologically irreducible. Of course non-holonomic constraints are theoretically reducible to microscopic dynamics, but this reduction obliberates their functional role. As \citet{neumann1955mathematical} demonstrated, if we describe both a physical system $S$ and its measuring device $M$ within a single dynamical framework, we then require a new device $M'$ to establish the initial conditions for the combined system $S + M$, and so on \emph{ad infinitum}. This infinite regress reveals a deeper issue: measurement---the act of determining the initial state of a system---cannot be internalized within the same dynamical description without erasing its function. 

A measurement device, in this context, is not merely another physical system but a structure that establishes the initial conditions under which dynamical laws operate. Measurement converts a physical state into a symbolic representation, and as such, it plays a role distinct from the dynamics it constrains. Attempting to model both the system and its measurement apparatus within a unified dynamic eliminates the crucial epistemic function of the measuring device as a selector and interpreter of system states. Consequently, measurement---and by extension, control (as the inverse process that translates symbolic instructions into physical outcomes)---cannot be fully captured within a closed dynamical description. The pragmatic function of measurement and control disappears when subsumed into purely energetic or dynamical laws. Thus, the epistemic distinction between symbol and referent, syntax and pragmatics, must be preserved if we are to retain explanatory power~\citep{cariani1989design}.

Such a distinction also underpins a critique towards computationalism. Turing’s abstract machine \citep{turing1936computable} manipulates symbols via syntactic rules independent of physical implementation. The illusion that symbol systems are autonomous (detached from time, energy, or semantic content) arises from this abstraction. Computation, while powerful for simulating complexity, cannot ground symbolic meaning in physical processes. This isolation is evident in programmable computers, where non-integrable hardware (e.g., switches, memory) allows symbolic control, yet the computer’s function appears implementation-free. As \citet{hoffmeyer1991code} and \citet{kull1998semiosis} note, symbolic expression seems unconstrained by physical law. Yet, this view is misleading: computation is independent of its material implementation only in ideal noiseless, error-free symbol manipulation~\citep{pattee2012evolving}; as \citet{rosen1987scope} and \citet{emmeche1996garden} argue, computationalism leads to syntacticalization and derealization, masking the physical substrate of symbolic functions. This insight suggests that capturing the concurrent, adaptive, and information-sensitive behavior observed in living systems~\citep{grozinger2019pathways} requires a departure from classical, Turing-style architectures. Instead, it invites the development of computing paradigms that explicitly incorporate mechanisms of measurement and control~\citep{calude1998unconventional}. 

In this way, based on the epistemic cut, Pattee partially answered Pearson's question: 
\emph{“It is not possible to distinguish the living from the lifeless by the most detailed “motion of inorganic corpuscles” alone. The logic of this answer is that life entails an epistemic cut that is not distinguishable by microscopic (corpuscular) laws. As von Neumann's argument shows, any distinction between subject and object requires a description of the constraints that execute measurement and control processes; and such a functional description is not reducible to the dynamics that is being measured or controlled.”}~\citep[p. 16]{pattee2001physics}. 

We can think of the genotype-phenotype interface as the primeval locus of this epistemic cut. Gene sequences are energetically degenerate and not distinguished by physical law; distinctions arise only when proteins fold into active configurations that break degeneracy and initiate function. These distinctions occur only within the context of a self-constructing system, making \emph{semantic closure}---the self-referential condition where a system’s symbols control the construction of its own interpretive mechanisms---necessary for life \citep{pattee2012cell, pattee2012evolving}. Semantic closure represents a circularity where genotype informs phenotype, which in turn constructs and interprets the genotype, bridging syntax and pragmatics through self-referential dynamics. However, as~\cite{walker2013algorithmic} emphasize, the real challenge of life’s origin is to understand how those instructional information control systems emerge naturally and spontaneously from mere molecular dynamics. Currently, we lack a general model to explain how such semantic closure was originated~\citep{pattee2012evolving}.

This paper explores the origin and early evolution of semantic closure. First, we introduce Rosen's thinking and analyze his (Metabolism, Repair)-systems, which later are extended to Hofmyer's (Fabrication, Assembly)-systems. Second, we develop a temporally parameterized extension of (F, A)-systems, unifying key properties of life, including autopoiesis, anticipation, and adaptation. Third, we elucidate the origin and early evolution of semantic closure by proceeding down the logical chain of assumptions, obtaining weaker versions of our model. Finally, we integrate these findings into a computational enactivist framework consistent with multiple theoretical results that have recently been obtained, showing that there is a self-referential takeover in the evolution of life. Throughout our manuscript, we provide a biologically relevant example of (M, R)-system, discuss the realizability of relational models, and propose some ways to extend our ideas, including open questions for our research program and a glossary explaining relevant terms that may be confusing to the reader.

\section{A Unified Representation of Life}
Throughout the last century, multiple models have been developed to explain the fundamental characteristics of life: the homeostat~\citep{ashby1949electronic}, (M, R)-systems~\citep{rosen1958relational, rosen1958representation}, the kinematic automaton~\citep{neumann1966theory}, autopoiesis~\citep{VarelaEtAl1974, maturana1980}, dissipative structures~\citep{prigogine1978time}, autocatalytic sets~\citep{kauffman1986autocatalytic, maley2025life}, the chemoton~\citep{ganti2003chemoton}, the hypercycle~\citep{eigen2012hypercycle}, among others. However, although the notion of organizational closure is ubiquitous in all these contemporary frameworks, remarkably none of them cited others~\citep{cornish2020contrasting}, leaving aside a potential unified model where all these conceptualizations turn out to be different faces of the same die. As~\citet{korbak2023self} noted, of all these models, Rosen’s (Metabolism, Repair)-systems are \textit{“general enough to encompass all kinds of mathematical structures (and formal or computational models),”} becoming an \emph{epistemological infrastructure} for all modeling in biology~\citep{varenne2013mathematical}.

In a long series of papers dating back to 1958 \citep{rosen1958relational, rosen1958representation}, and later condensed in the 1990s~\citep{rosen1991life}, Robert Rosen introduced the (Metabolism, Repair)-systems paradigm, which was inspired by biotopology~\citep{rashevsky1954topology}. At first, Rosen represented (M, R)-systems using graph theory, but he quickly transitioned to category theory, as this shift allowed him to better represent the minimal functional relationships required for life. Historically,  Rosen's early papers predate the widespread use of the Yoneda lemma, the Lawvere's fixed-point theorem, and other fundamental category theory results. However, later reconstructions showed (M, R)-systems can indeed be derived functorially via Yoneda~\citep{baianu1974functorial}.

Instead, Rosen’s adoption of category theory followed Eilenberg-MacLane’s original viewpoint: any category embeds faithfully into \textbf{Set} (the category of sets), so one may reason with sets without forfeiting structural generality~\citep{eilenberg1945general, rosen1958representation}. Our aim here is expository and biological: we keep the categorical language to make causes, closure, and realizability model-independent, but we avoid imposing extra axioms that would be needed to deploy powerful theorems that are not necessary for the present results. 

Let $\mathcal{C}$ be a \gls{cc}, $A, B \in \text{Ob}(\mathcal{C})$, $f \in H(A, B)$, and $\Phi_f\in H(B, H(A, B))$, where $H(X, Y)$ is a proper subset of the hom-set of all mappings from $X$ to $Y$. According to~\citet{rosen1971some}, by making $A$, $B$, and $f$ sufficiently complex, it is possible to reduce every (M, R)-system to the following diagram:

\begin{equation}
    A \xrightarrow{f} B \xrightarrow{\Phi_f} H(A, B).
    \label{eq:simplest}
\end{equation}

Unlike the traditional Newtonian approach, which generally uses state-space-based models with well-defined transition functions, Rosen’s approach relies on temporally invariant category theory models, which support multiple modes of causation consistent with Aristotle's four causes. According to Rosen, it is this very segregation into independent categories of causation that prevents the Newtonian picture from describing the entailments and linkage relations characteristic of life~\citep{rosen1978anticipatory, rosen1991life}. In diagram~(\ref{eq:simplest}), $A$ is the \gls{material} of $B$, $f$ is the \gls{efficient} of $B$, and $H(A, B)$ is the \gls{final} of $B$. Here, Rosen leaves out the \gls{formal} while making the efficient cause of $f$ (which is $\Phi_f$) internal to the system. At the same time, the efficient cause of $\Phi_f$ lies outside the system, requiring an external agent. To resolve this issue, Rosen used the evaluation map of $\mathcal{C}$, which is defined as follows:

\begin{equation}
    \text{eval}_{Y, X}^{\mathcal{C}}: Y^X\times X \xrightarrow{} Y,
    \label{eq:evaluation}
\end{equation}

where $Y^X$ denotes the hom-set of all mappings from $X$ to $Y$. The evaluation map $\text{eval}_{Y,X}^{\mathcal{C}}$ takes a mapping $f\in Y^X$, an input $x\in X$, and returns the value $f(x)\in Y$. Importantly, given $f \in Y^X$, if we fix a specific $x \in X$ and vary $f \in Y^X$, we can define a new mapping $\hat{x}: Y^X \xrightarrow{} Y$, which satisfies $f(x) = \hat{x}(f)$. For Rosen's purposes, the \emph{inverse} evaluation map allows the construction of hierarchical cycles that underlie his (M, R)-systems. Building on this foundation, Rosen extended the framework to explicitly incorporate closure. By letting $X = B$ and $Y = H(A, B)$ in Eq.~(\ref{eq:evaluation}), we obtain:

\begin{equation*}
    \hat{b}: H(B, H(A, B))\xrightarrow{} H(A, B).
\end{equation*}

Mapping $\hat{b}$ has a left-inverse if it is injective. This means that for every pair of maps $\Phi_1,\Phi_2\in H(B, H(A, B))$, 

\begin{equation*}
    \hat{b}(\Phi_1) = \hat{b}(\Phi_2)\Rightarrow\Phi_1 = \Phi_2.
\end{equation*}

Equivalently,

\begin{equation}
\Phi_1(b) = \Phi_2(b)\Rightarrow\Phi_1=\Phi_2.
\label{eq:inj_rep}
\end{equation}

Implication~(\ref{eq:inj_rep}) tells us that if two replacement maps agree at $b$, then they must agree everywhere. Thus, the replacement map $\Phi_f$ is uniquely determined by its one value $\Phi_f(b) = f\in H(A,B)$, endowing closure in (M, R)-systems with nothing but the ingredients of metabolism and replacement. Let us denote $\hat{b}^{-1}$ as $\beta_b$. Through this construction, Rosen derived a new (M, R)-system that incorporates closure, expressed as:

\begin{equation}
   B \xrightarrow{\Phi_f} H(A, B) \xrightarrow{\beta_b} H\left(B, H(A, B)\right).
    \label{eq:simplest2}
\end{equation}

In Rosen's words: \textit{“This construction has two important consequences: (a) it is the only formalism, to my knowledge, which incorporates replication of genetic components as an essential and natural consequence of metabolic activities, and (b) exhibits a new and most interesting formal equivalence between metabolic and replicative aspects in these systems”}~\citep{rosen1971some}. By combining the diagrams~(\ref{eq:simplest}) and~(\ref{eq:simplest2}), Rosen arrived at the canonical (M, R)-system:

\begin{equation}
   \beta_b\circ \Phi_f\circ f: A \xrightarrow{f} B \xrightarrow{\Phi_f} H(A, B) \xrightarrow{\beta_b} H\left(B, H(A, B)\right).
    \label{eq:classic}
\end{equation}

In~\citet{rosen1971some}'s terminology: $f$ is the metabolic part (M) of the system~(\ref{eq:classic}), $\Phi_f$ is the repair (R) part of the system~(\ref{eq:classic}) and $\beta_b$ is the replicative part of the system~(\ref{eq:classic}). As~\citet{letelier2011homme} explains, it is much more logical from a biological point of view to use the terms “replacement” and “closure” rather than “repair” and “replication.” \citet{rosen1991life} expresses~(\ref{eq:classic}) as the following diagram:

\begin{equation}
\label{d:rosen}
\begin{tikzcd}
  & f \arrow[dl, color=black, dashed, thick] \arrow[dr, color=black, thick]  &  \\
A \arrow[r, color=black, thick] & \arrow[u, color=black, thick, transform canvas={xshift=-0.9ex}] B \arrow[u, color=black, dashed, thick, transform canvas={xshift=0.9ex}] & \arrow[l, color=black, dashed, thick] \Phi_f   
\end{tikzcd}
\end{equation}

In diagram~(\ref{d:rosen}) the solid arrows represent the mappings showed in~(\ref{eq:classic}) and the dashed arrows represent the morphisms $f$, $\Phi_f$, $\beta_b$ being applied to their respective input. As the reader could observe, implicitly Rosen is assuming there is a one-to-one correspondence between $B$ and $\beta_b$~\citep{cardenas2010closure}. This guarantees \textit{closure to efficient causation}. In biological terms, this means that all enzymes, all catalytic processes that are used within the system, need to be synthesized by the organization \emph{per se}, without recourse to any external agent. According to Rosen, this is what characterizes life~\citep{rosen1991life}. 

\subsection{A biological example}

In the origins-of-life literature, reflexively autocatalytic and food-generated (RAF) networks~\citep{letelier2010m} and their seed-dependent autocatalytic system (SDAS) variants~\citep{cornish2020contrasting} give concrete instantiations of organizational closure at a chemical level. In RAF/SDAS, a fixed “food set” together with catalysis constraints selects admissible reaction maps. Seeds trigger higher-level closures that recursively generate their own catalysts. This clearly mirrors the role of $H(A, B)$ (admissible metabolisms) and $\Phi_f$ (selector of the executable metabolism) here, while the replication step $\beta_b$ corresponds to the passage to higher-order closures once a viable catalytic configuration $b$ is present. Thus, our categorical (M, R)-scaffold offers a unifying language that can explicitly host RAF/SDAS chemistries as particular choices of objects, morphisms, and admissibility constraints~\citep{letelier2006organizational}.

So far, our treatment of characterizing life has been heavily abstract. To help the reader digest this content, and at the same time, provide a clearer perspective on relational biology, we dedicate this short section to exemplify how can we map Rosen’s abstraction to biologically-relevant morphisms and objects. First, let us coarse-grain the core gene-expression machinery of a bacterial cell into four catalysts:

\[
\text{RNAP},~ \text{Rib},~ \text{AARS},~\text{Chap},
\]

standing respectively for RNA polymerase, a ribosome, the (lumped) set of aminoacyl-tRNA synthetases, and a chaperone/assembly factor. \textcolor{blue}{As \emph{materials} (transformables), we consider pools of precursors:}

\[
\begin{gathered}
\mathrm{AA}\ (\text{amino acids}),\ \mathrm{NTP}\ (\text{nucleoside triphosphates}),\\
\mathrm{tRNA},\ \mathrm{rRNA},\ \textcolor{blue}{\mathrm{mRNA}}.
\end{gathered}
\]

\textcolor{blue}{We treat the genome $G$ as a fixed parameter constraining admissible maps. The objects are now specified as \emph{sets of admissible configurations} and their cartesian products:}

\[
\textcolor{blue}{A\;:=\; \mathrm{AA}\times \mathrm{NTP}\times \mathrm{tRNA}\times \mathrm{rRNA}\times \mathrm{mRNA},}
\]

and

\[
\textcolor{blue}{B\;:=\; \mathrm{RNAP}\times \mathrm{Rib}\times \mathrm{AARS}\times \mathrm{Chap}.}
\]

\textcolor{blue}{Therefore, an element $a\in A$ gives material levels/compositions, while $b\in B$ is a 4-tuple of catalytic states. With this, metabolism $f\in H(A,B)$ is the composite action of the following schematic reactions (arrows labeled by their catalysts):}

\[
\resizebox{\columnwidth}{!}{$
\begin{aligned}
\textbf{Transcription:}\quad
& \mathrm{NTP} \xrightarrow{\ \mathrm{RNAP}\ } \mathrm{mRNA}_{\mathrm{rpo}},\ \mathrm{mRNA}_{\mathrm{ribo}},\ \mathrm{mRNA}_{\mathrm{aars}},\ \mathrm{rRNA},\ \mathrm{tRNA},\\
\textbf{Charging:}\quad
& \mathrm{AA}+\mathrm{tRNA} \xrightarrow{\ \mathrm{AARS}\ } \mathrm{aa\mbox{-}tRNA},\\
\textbf{Translation:}\quad
& \mathrm{aa\mbox{-}tRNA}+\mathrm{mRNA} \xrightarrow{\ \mathrm{Rib}\ } P \\
\textbf{Folding/assembly:}\quad
& P+\mathrm{rRNA} \xrightarrow{\ \mathrm{Chap}\ } \mathrm{RNAP},\ \mathrm{Rib},\ \mathrm{AARS},\ \mathrm{Chap}.
\end{aligned}
$}
\]

\textcolor{blue}{Where $P$ represents the pool of unfolded polypeptides, including RNAP/Rib/AARS/Chap chains. At the level of $\mathcal C$, this induces a time-invariant morphism}

\[
f:\; A \longrightarrow B,~~ a\longmapsto b=f(a),
\]

Let \(H(A,B)\) be the set of \emph{admissible} metabolic maps compatible with (i) the fixed genome \(G\), (ii) mass-balance/assembly constraints, and (iii) coarse-grained kinetics. \textcolor{blue}{Given a catalytic configuration $b\in B$ (current RNAP/Rib/AARS/Chap states), the \emph{selector/replacement} map}

\[
\Phi_f\in H\bigl(B,\,H(A,B)\bigr)
\]

returns the unique admissible metabolism \(f_b\in H(A,B)\) that \emph{this very configuration} can realize on genome \(G\):

\[
\textcolor{blue}{\Phi_f(b) \;=\; f_b,~~ \text{with}~~ f_b(a)=b'\in B.}
\]

\textcolor{blue}{Intuitively, the particular RNAP holoenzyme, ribosome stoichiometry, and AARS repertoire determine which $A\!\to\!B$ map is executable (promoter usage, translation capacity, assembly constraints).} Now, let us pick a \emph{distinguished} self-consistent configuration

\[
\textcolor{blue}{b^\star=\bigl(\mathrm{RNAP}_{\sigma},\ \mathrm{Rib}_{70\mathrm{S}},\ \mathrm{AARS}_{\mathrm{complete}},\ \mathrm{Chap}_{\ast}\bigr)\in B,}
\]

\textcolor{blue}{where $\mathrm{RNAP}_{\sigma}\in\mathrm{RNAP}$ is a housekeeping $\sigma$-holoenzyme, $\mathrm{Rib}_{70\mathrm{S}}\in\mathrm{Rib}$ an assembled ribosome, $\mathrm{AARS}_{\mathrm{complete}}\in\mathrm{AARS}$ a complete minimal set, and $\mathrm{Chap}_{\ast}\in\mathrm{Chap}$ an adequate chaperone pool. We can restrict the selector space to those admissible under $G$,}

\[
S(G)\;:=\;H\bigl(B,\,H(A,B)\bigr).\
\]

$S(G)$ contains selectors that (a) read the same genome \(G\), (b) preserve the assembly/stoichiometry rules, and (c) differ, if at all, only by regulation consistent with \(b^\star\). \textcolor{blue}{On this reduced $S(G)$ we assume an injective evaluation at $b^\star$:}

\[
\forall\,\Phi_1,\Phi_2\in S(G): \Phi_1(b^\star)=\Phi_2(b^\star)\ \Longrightarrow\ \Phi_1=\Phi_2.
\]

Equivalently, the evaluation map

\[
\text{eval}_{A, B}^{\mathcal{C}}:\ S(G)\longrightarrow H(A,B),~~ \Phi\longmapsto \Phi(b^\star),
\]

is one-to-one. \textcolor{blue}{Hence it admits a left inverse on its image,}

\[
\beta_{b^\star}\ :=\ \text{eval}_{Y, X}^{\mathcal{C}~~~-1}:\ H(A,B)\longrightarrow S(G)\ \subseteq\ H\bigl(B,H(A,B)\bigr),
\]

\textcolor{blue}{so that $\beta_{b^\star}(f)=\Phi_f$ and $\Phi_f(b^\star)=f$. In this way, the $(M,R)$-loop is realized by the gene-expression core under $G$, with \emph{objects} of $\mathcal C$ read as material/catalyst pools and \emph{morphisms} as genome- and assembly-constrained transformations between them. This same reading also lets RAF/SDAS early metabolisms serve as worked examples: $A$ is a food-plus-intermediate pool, $B$ the active catalyst set, $H(A,B)$ the reaction-closure maps permitted by catalysis and availability; $\Phi_f$ selects which closure a given $b\in B$ can realize, and $\beta_b$ records how one viable configuration pins down the entire selector.}

\subsection{Realizing relational models}
Rosen’s assertion that no model of closure to efficient causation could be Turing-simulable has often been misunderstood or misrepresented in the literature~\citep{lane2024robert}. As emphasized by \citet{cornish2020contrasting}, many critiques of Rosen's claim stem from a failure to grasp the core logic of relational biology. This is entirely understandable, as it is a fact that Rosen's ideas changed abruptly over time~\citep{pattee2007laws}. For example, Rosen himself investigated the representation of relations, such as~(\ref{eq:simplest}), using sequential machines, extensively exploring the implications of this approach \citep{rosen1964abstract, rosen1964abstract2, rosen1966abstract}. It was the inclusion of replication mechanisms, such as those represented by~(\ref{eq:simplest2}), that posed challenges ~\citep{baianu1974functorial}. 

Later, in what is possibly his best known work~\citep{rosen1991life}, he never used the term \textit{computable}. Instead, Rosen referred to machines, mechanisms, and the notion of simulation. Nevertheless, we can get an idea of his late notion of computability from there: 

\textit{“Thus, the word ‘simulable’ becomes synonymous with ‘evaluable by a Turing machine’. In the picturesque language of Turing machines, this means the following: if $f$ is simulable, then there is a Turing machine $T$ such that, for any word $w$ in the domain of $f$, suitably inscribed on an input tape to $T$, and for a suitably chosen initial state of~~$T$, the machine will halt after a finite number of steps, with $f(w)$ on its output tape.”} \citep[p. 192]{rosen1991life}

From the above, a myriad of interpretations have arisen, even calling into question the validity of Rosen's arguments~\citep{cardenas2010closure}. Despite this, it is a fact that Rosen's reasoning lies in the impossibility of capturing self-reference by means of sequential models of computation. In the same way that a Turing machine cannot decide whether it halts on its own description (the Halting Problem), a single automaton cannot decide its own entailment structure when the entailments depend on itself in a circular manner. For example, if $f$ is produced by $B$, but $f$ is also the catalyst that turns $A$ into $B$, then we end up needing $f$ to define $f$. This is a classic impredicative loop~\citep{soto2011ouroboros}. Thus, even if we allow dynamic label reassignment, that does not help with the causal loop unless the automaton can change its own state transition structure during computation based on its own internal outputs. That would require meta-computation, something not allowed in classical computer science.

Using $\lambda$-calculus to represent relational models is theoretically appealing~\citep{mossio2009computable}. Nevertheless, as shown by~\citet{cardenas2010closure}, the mathematical solutions it provides may not directly correspond to biological phenomena, necessitating further refinement and interpretation. A promising resolution to this challenge is provided by \citet{palmer2016rosen}, who propose that while a single automaton cannot model an organism, a set of automata communicating and sharing information might achieve this. It is clear that~\citet{palmer2016rosen} wanted to create a \emph{model} of Rosen's replicative (M, R)-system rather than just another simulation~\citep{mossio2009computable, gatherer2013rosen, zhang2016rosen}. However, at the end of their manuscript, they clarified that their communicating X-machine model does not, as the modeling relation requires, map to the (M, R)-system relations~\citep{palmer2016rosen}. This creates even more obstacles if we want to \emph{model} (not simulate) fabrication, self-assembly, and an actively maintained assembly-enabling environment. As we shall see later, these three aspects are essential to expand Rosen's ideas.

As~\citet{pattee2019simulations} points out, simulation accuracy does not translate into effective functional realizations. Most importantly, simulations should never be conflated with realizations. When we simulate life we get rid of its semantic and pragmatic aspects, focusing merely on its syntactic nature. Simulations---and more generally, consistent formal systems (i.e., not allowing contradictions)---require a predefined set of rules in which novelty is limited. Realizations of life, on the other hand, are capable of generating novelty by interacting directly with their environment, capturing with it those semantic-pragmatic aspects that simulation lacks. This distinction is critical because it underscores the limitations of purely algorithmic models in fully capturing the self-referential and anticipatory nature inherent in biological systems~\citep{jaeger2024naturalizing}. We have characterized computation as transformations from symbols to symbols, emphasizing that both measurement and control are fundamentally irreducible to computation \citep{neumann1955mathematical}. Thus, if instead of merely simulating, we construct realizations of relational models, we can inherently incorporate measurement and control into their architecture, becoming capable of interacting with and adapting to their environment. This blueprint effectively transcends the purely symbolic definition of computation we initially presented and allows us to better capture the essence of life~\citep{cariani1989design}.

Another relevant aspect when building life realizations is its inherent stochasticity~\citep{grozinger2019pathways}. At the molecular origin of life, stochasticity manifested significantly through ambiguity in early genetic codes and high translational error rates, resulting in heterogeneous peptide populations, called \emph{statistical proteins}~\citep{barbieri2018code}. These statistical proteins acted as transitional molecular forms that allowed primitive biological systems to broadly explore chemical and functional spaces, potentially providing adaptive advantages and evolutionary pathways toward increased molecular specificity and coding accuracy. Therefore, regardless of its implementation substrate, any realization of life must be intrinsically stochastic.

\subsection{(F, A)-systems}

(M, R)-systems face two significant challenges. First, they depend on injectivity in $H(A, B)$ to ensure closure, a condition that imposes strict mathematical constraints for the concrete category $\mathcal{C}$~\citep{rosen1963some, letelier2006organizational}. Second, the one-gene-one-enzyme hypothesis that originally inspired Rosen’s model is now recognized as overly simplistic~\citep{horowitz1948one}. Thus, Rosen's model is not biologically sound. A promising alternative that addresses these limitations is provided by (Fabrication, Assembly)-systems~\citep{hofmeyr2021biochemically}. These systems extend Rosen’s approach by incorporating freestanding formal causes, allowing for the modeling of both physiological and evolutionary adaptation.

To proceed, it is necessary to clarify what we mean by formal cause. As \citet{hofmeyr2018causation} explains, formal causation can be integrated with Aristotle’s other causes (material and efficient) in multiple ways. Broadly, formal causes can be either freestanding or intrinsic. In the freestanding case, the formal cause exists as an independently specified sequence of rules that directs either the material or efficient cause in carrying out a transformation toward a final cause. Examples include a catalyst-template system where monomers are assembled into polymers, or tagging enzymes that act on target-tag pairs to produce a tagged target. In the first example, the enzyme (efficient cause) works in tandem with a separate template (formal cause); in the second, the material cause (target) interacts with an independent tag acting as the formal cause.

By contrast, intrinsic formal causation occurs when the formal cause is embedded within the structure or properties of the material or efficient cause itself. Examples include enzyme active sites with specific substrate-binding properties, and unfolded polymers folding into functional conformations under suitable conditions. In the first case, the enzyme active site is an informed efficient cause---it can act only on specific substrates. In the second, the monomer sequence of the polymer acts as an informed material cause---it folds predictably when placed in a particular physical context. After this clarification, we can now show the simplest representation of an (F, A)-system:

\begin{align}
    A \xrightarrow{f} B \xrightarrow{\Phi_f} H(A, B), \label{eq:FA-1} \\
    D \xrightarrow{f} H\left(B, H(A, B)\right), \label{eq:FA-2} 
\end{align}

where $f=I\times\{f_1\}+f_2$, $I=\{I_1, I_2\}$, and $B=B_1+B_2$. Here, the addition symbol is not a simple algebraic sum but rather indicates that $f_2$ is being incorporated into the overall process. In (F, A)-systems, $f$ has a dual role: $f_1$, guided by freestanding formal causes ($I$), produces $B$, while $f_2$, an informed efficient cause, produces $\Phi_f$. This allows the system to remain closed to efficient causation while being open to formal causation~\citep{hofmeyr2018causation}. 

From a biosemiotic perspective, this structure becomes especially revealing. As~\citet{cariani1989design} points out, three of Aristotle’s causes align naturally with the semiotic triad identified by~\citet{morris1946signs}: syntax corresponds to freestanding formal cause, semantics to efficient cause, and pragmatics to final cause. Furthermore, material cause translates as symbol vehicle (substrate), informed efficient causes are semantic interpreters whose specificity is embedded within them, and informed material causes are symbolic substrates whose syntax is materially inscribed. This translation is highly suggestive. If (F, A)-systems are closed to efficient causation but open to freestanding formal causation, they fundamentally exhibit semantic closure and possess adaptive syntax. The latter implies that the system is no longer constrained by a fixed rule set, thus enabling the representation of \glspl{affordance}~\citep{kauffman2021world}. 

As a result, (F, A)-systems achieve organizational closure and heritable variability while circumventing the need for injectivity in $H(A, B)$ as a prerequisite for closure, thereby rendering the model more biologically realistic. Perhaps a much more intuitive way to understand the above is to look at the diagram that represents this class of models:

\begin{equation}
\label{d:hofmeyr}
\begin{tikzcd}
  & \textcolor{violet}{I}\times\{\textcolor{magenta}{f_1}\}+\textcolor{blue}{f_2} \arrow[dl, color=magenta, dashed, thick] \arrow[r, color=blue, dashed, thick]  & D \arrow[d, color=blue, thick] \\
A \arrow[r, color=magenta, thick] & \arrow[u, color=orange, thick] B_1+B_2 & \arrow[l, color=orange, dashed, thick] \textcolor{orange}{\Phi_f}   
\end{tikzcd}
\end{equation}

As shown by~\cite{hofmeyr2021biochemically}, (F, A)-systems allow the interpretation of concepts such as the metabolism and repair components of Rosen’s (M, R)-system~\citep{rosen1958representation}, John von Neumann’s kinematic automaton~\citep{neumann1966theory}, Howard Pattee’s symbol-function split via the symbol-folding transformation~\citep{pattee2012clues}, Marcello Barbieri’s genotype–ribotype–phenotype ontology~\citep{barbieri1981ribotype}, and Tibor Gánti’s chemoton~\citep{ganti2003chemoton}. Thus, (F, A)-systems are endowed with features that characterized life, such as thermodynamic openness, catalytic specificity, catalytic closure, structural closure, information handling, reproduction, and evolvability~\citep{cornish2020contrasting}. 

In this way, (F, A)-systems provide a canonical framework for representing biological organization, demonstrating how hierarchical cycles cannot be fully captured by algorithmic models due to their intrinsic collective \gls{impredicativity}~\citep{jaeger2024naturalizing}. For further details on how (F, A)-systems can represent the logic underlying a self-manufacturing cell, we recommend consulting~\citet{hofmeyr2021biochemically}. In the next section, we demonstrate how (F, A)-systems can be temporally extended, naturally giving rise to anticipatory systems. This integration allows us to build a computational enactivist narrative that resolves the limitations of algorithmic approaches to modeling anticipation in non-human and non-cognitive living systems.

\section{Extending (F, A)-Systems}
One limitation of the current (Fabrication, Assembly)-system model is its blindness to the external environment. While it accounts for internal self-maintenance, it does not explicitly incorporate mechanisms for environmental sensing and response, both of which are essential for biological measurement (represented by perception carried out by sense organs) and control (represented by actions carried out by various effectors, such as motor and secretory organs)~\citep{cariani1989design}. Notwithstanding, Hofmeyr notes that this limitation can be addressed by incorporating membrane receptors and signal transduction networks into the fabrication machinery~\citep{hofmeyr2021biochemically}. Because these components are encoded in DNA, they can be naturally integrated into the (F, A)-system manufacturing architecture. This insight highlights the importance of extending the model to capture adaptive responses, further enhancing its biological realism from an ecological point of view.

Let $I=\{I_1, I_2, I_3, I_4\}$ be an extended set of freestanding formal causes, where $I_3$ and $I_4$ specifically represent the instructions for manufacturing the sensing and response architectures, respectively. Associated respectively with $I_3$ and $I_4$, we have $f_3:X\to Y$ (membrane receptors) and $f_4:Y\to Z$ (signal transducers) for external signals $X$, internal signals $Y$, and intracellular responses $Z$. These mappings are such that

\begin{equation*}
    X \xrightarrow{f_3} Y \xrightarrow{f_4} Z
\end{equation*}

Notice that these mappings do not form part of covalent metabolic chemistry. Thus, we can extend diagram~(\ref{d:hofmeyr}) as follows:

\begin{equation}
\label{d:hofmeyr-ext}
\begin{tikzcd}[column sep=0 cm]
& X \ar[r,thick] & Y \ar[r,thick] & Z \\
&& f = I \times \{f_1\}+f_3+f_4+f_2 \ar[lld,dashed,thick,start anchor={[xshift=-1ex]}]
\ar[rr,dashed,thick] \ar[ul,dashed,thick,start anchor={[xshift=6ex]}]
\ar[u,dashed,thick,start anchor={[xshift=6ex]}] & & D \ar[d,thick]   \\
A \ar[rr,thick] & & B_1+B_2+B_3+B_4 \ar[u,thick] & & \Phi \ar[ll,thick]
\end{tikzcd}
\end{equation}

In diagram~(\ref{d:hofmeyr-ext}) $B_3$ and $B_4$ are the components from which membrane transporters and signal transducers are assembled (their unfolded polypeptides). The use of freestanding formal causes ($I_3$, $I_4$) to encode membrane receptors and signaling networks opens a door to embedding environmental affordances~\citep{gibson1979ecological}. In this context, affordances are not fixed properties of the world but \emph{transjective}, relational possibilities for action that emerge through an organism’s embodied engagement with its environment. This perspective, emphasized in both ecological psychology~\citep{gibson1979ecological} and evolutionary accounts of agency~\citep{jaeger2024naturalizing}, highlights the role of the agent in delimiting a meaningful “arena” of possible interactions. Similar to the view defended in~\citet{kauffman2021world}, affordances are neither reducible to observer-independent features nor internal subjective projections; rather, they are \textit{enacted} through the dynamic, co-constitutive relation between organism and environment. Incorporating this framework within the extended (F, A)-system highlights its capacity not just for self-maintenance but also for situated adaptive behavior.

Biological systems, however, adapt across multiple time scales, including the rapid biochemical reactions of molecular processes, short-term learning, development, and evolution. These temporal scales are critical to understand how short-term dynamics influence long-term stability and vice versa~\citep{montevil2015biological}. To capture this interplay, we associate distinct time scales to each process within the (F, A)-system. Importantly, this approach aligns with the notion of constraint developed by~\citet{montevil2015biological}, which is equivalent to Rosen's concept of efficient causation. We start by rewriting the fundamental relations in our extended (F, A)-systems:

\begin{align} \label{eq:FA-explicit}
    &A \xrightarrow{f} B \xrightarrow{\Phi_f} H(A, B),\\ \nonumber 
    &D \xrightarrow{f} H\left(B, H(A, B)\right),\\ \nonumber
    &X \xrightarrow{f_3} Y \xrightarrow{f_4} Z,
\end{align}

where $f=I\times\{f_1\}+f_2+f_3+f_4$, $I=\{I_1, I_2, I_3, I_4\}$, and $B=B_1+B_2+B_3+B_4$. Each arrow shown in~(\ref{eq:FA-explicit}) represents a process, and thus it will have a characteristic temporal scale. For simplicity, let $C = H(A, B)$ and $E = H\left(B, H(A, B)\right)$. To emphasize the multifunctional role of $f$, we can rename the different mappings of~(\ref{eq:FA-explicit}) in the following way:

\begin{align} \label{eq:FA-explicit-simp}
    &A \xrightarrow{\phi} B \xrightarrow{\psi} C\\ \nonumber 
    &D \xrightarrow{\theta} E\\ \nonumber
    &X \xrightarrow{\alpha} Y \xrightarrow{\beta} Z.
\end{align}

Thus, we assign a time scale $\tau_{x}$ to each process $x\in\{\phi, \psi, \theta, \alpha, \beta\}$. The mappings in~\eqref{eq:FA-explicit-simp} can now be rewritten as:

\begin{align} \label{eq:FA}
    &A \xrightarrow[\tau_{\phi}]{\phi} B \xrightarrow[\tau_{\psi}]{\psi} C\\ \nonumber 
    &D \xrightarrow[\tau_{\theta}]{\theta} E\\ \nonumber
    &X \xrightarrow[\tau_{\alpha}]{\alpha} Y \xrightarrow[\tau_{\beta}]{\beta} Z.
\end{align}

Following~\citet{montevil2015biological}, two scales must be considered for every efficient causation included in a closed system: one scale at which the efficient causation is associated with a time symmetry and another at which it is produced and/or maintained. Let $\Delta_{\phi}=\tau_{\phi}-\tau_{\psi}$, $\Delta_{\psi}=\tau_{\psi}-\tau_{\theta}$, $\Delta_{\theta}=\tau_{\theta}-\tau_{\psi}$, $\Delta_{\alpha}=\tau_{\alpha}-\tau_{\psi}$, and $\Delta_{\beta}=\tau_{\beta}-\tau_{\psi}$. Since the system is closed to efficient causation, at least one $\Delta_i$ must be positive, and another $\Delta_j$ ($j\neq i$) must be negative~\citep{montevil2015biological}. This maintains the temporal coherence of the organization, allowing for organizational closure as the system unfolds over time. Using the methodology proposed by~\citet{korbak2023self}, we represent the temporal parametrization of~\eqref{eq:FA} as the asynchronous dynamic Bayesian network shown in Figure~\ref{fig:ADBN}.

\begin{figure*}[ht!]
\centering
\begin{adjustbox}{max width=\textwidth}
\begin{tikzcd}[column sep=5.5em, row sep=40ex, >={Latex}, thick]
A_t \arrow[r,"\phi_t"] & B_t \arrow[r,"\psi_t"] & C_t &
D_t \arrow[r,"\theta_t"] & E_t &
X_t \arrow[r,"\alpha_t"] & Y_t \arrow[r,"\beta_t"] & Z_t \\
A_{t+\tau} \arrow[r,"\phi_{t+\tau}"{name=phi_ttau}] &
B_{t+\tau} \arrow[r,"\psi_{t+\tau}"{name=psi_ttau}] & C_{t+\tau} &
D_{t+\tau} \arrow[r,"\theta_{t+\tau}"{name=theta_ttau}] & E_{t+\tau} &
X_{t+\tau} \arrow[r,"\alpha_{t+\tau}"{name=alpha_ttau}] &
Y_{t+\tau} \arrow[r,"\beta_{t+\tau}"{name=beta_ttau}] & Z_{t+\tau}
%
\arrow[from=1-3, to=phi_ttau,   bend right=18, dashed, shorten <=2pt, shorten >=2pt, "\tau_{\phi}"']
\arrow[from=1-3, to=theta_ttau, bend right=38, dashed, shorten <=2pt, shorten >=2pt, "\tau_{\theta}"]
\arrow[from=1-5, to=psi_ttau,   bend right=10, dashed, shorten <=2pt, shorten >=2pt, "\tau_{\psi}"']
\arrow[from=1-3, to=alpha_ttau, bend right=20,  dashed, shorten <=2pt, shorten >=2pt, "\tau_{\alpha}"]
\arrow[from=1-3, to=beta_ttau,  bend left=20,  dashed, shorten <=2pt, shorten >=2pt, "\tau_{\beta}"']
\end{tikzcd}
\end{adjustbox}
\caption{Temporal parametrization of extended (F, A)-systems.}
\label{fig:ADBN}
\end{figure*}

\begin{figure*}[ht!]
\centering
\begin{adjustbox}{width=\textwidth}
\begin{tikzcd}[column sep=3.2em, row sep=32ex, >={Latex}, thick]
A_t \arrow[r] & \phi_t \arrow[r] & B_t \arrow[r] & \psi_t \arrow[r] & C_t &
D_t \arrow[r] & \theta_t \arrow[r] & E_t &
X_t \arrow[r] & \alpha_t \arrow[r] & Y_t \arrow[r] & \beta_t \arrow[r] & Z_t \\
A_{t+\tau} \arrow[r] & \phi_{t+\tau} \arrow[r] & B_{t+\tau} \arrow[r] & \psi_{t+\tau} \arrow[r] & C_{t+\tau} &
D_{t+\tau} \arrow[r] & \theta_{t+\tau} \arrow[r] & E_{t+\tau} &
X_{t+\tau} \arrow[r] & \alpha_{t+\tau} \arrow[r] & Y_{t+\tau} \arrow[r] & \beta_{t+\tau} \arrow[r] & Z_{t+\tau}
%
\arrow[from=1-5, bend right=12, to=2-2,  shorten <=2pt, shorten >=2pt, "\tau_{\phi}"]
\arrow[from=1-5, bend right=22, to=2-7,  shorten <=2pt, shorten >=2pt, "\tau_{\theta}"]
\arrow[from=1-8, bend right=12, to=2-4,  shorten <=2pt, shorten >=2pt, "\tau_{\psi}"]
\arrow[from=1-5, bend right=12,  to=2-10, shorten <=2pt, shorten >=2pt, "\tau_{\alpha}"]
\arrow[from=1-5, bend left=12,  to=2-12, shorten <=2pt, shorten >=2pt, "\tau_{\beta}"]
\end{tikzcd}
\end{adjustbox}
\caption{Probabilistic Graphical Model associated to the temporal parametrization of extended (F, A)-systems.}
\label{fig:PGM}
\end{figure*}

In Fig.~\ref{fig:ADBN}, $\tau = \tau\left(\tau_{\phi}, \tau_{\psi}, \tau_{\theta}, \tau_{\alpha}, \tau_{\beta}\right)$, allowing us to model multiple parallel dependencies and interactions probabilistically. Following~\citet{korbak2023self}, we reinterpret each component and each solid line as a random variable, while reinterpreting each dashed line as a dependency. The corresponding probabilistic graphical model is shown in Figure~\ref{fig:PGM}. Thus, the simulation of an (F, A)-system reduces to inferring the following joint distribution:

\begin{equation}
\resizebox{\columnwidth}{!}{$
\begin{aligned}
p(&\phi_{\tau}, \psi_{\tau}, \theta_{\tau}, \alpha_{\tau}, \beta_{\tau},
    A_{\tau}, B_{\tau}, C_{\tau}, D_{\tau}, E_{\tau}, X_{\tau}, Y_{\tau}, Z_{\tau}, \ldots, \\
  & \phi_{T}, \psi_{T}, \theta_{T}, \alpha_{T}, \beta_{T}, A_{T}, B_{T}, C_{T}, D_{T}, E_{T}, X_{T}, Y_{T}, Z_{T}) =
\\
&\prod_{t \ge 0}^{T-\tau} \Big[
    p(A_{t+\tau})
    \, p(\phi_{t+\tau} \mid A_{t+\tau}, C_t)
    \, p(B_{t+\tau} \mid \phi_{t+\tau})
    \, p(\psi_{t+\tau} \mid B_{t+\tau}, E_t)
\\
&\qquad\quad
    p(C_{t+\tau} \mid \psi_{t+\tau})
    \, p(D_{t+\tau})
    \, p(\theta_{t+\tau} \mid D_{t+\tau}, C_t)
    \, p(E_{t+\tau} \mid \theta_{t+\tau})
    \, p(X_{t+\tau})
\\
&\qquad\quad
    p(\alpha_{t+\tau} \mid X_{t+\tau}, C_t)
    \, p(Y_{t+\tau} \mid \alpha_{t+\tau})
    \, p(\beta_{t+\tau} \mid Y_{t+\tau}, C_t)
    \, p(Z_{t+\tau} \mid \beta_{t+\tau}) \Big].
\end{aligned}
$}
\end{equation}

Through ancestral sampling, we can infer this distribution~\citep{korbak2023self}. Moreover, this temporal parametrization links (F, A)-systems to active inference, conceptualizing primitive living systems as generative models probabilistically tailoring their internal states and environments. Since the concept of Markov blanket~\citep{friston2013life} is compatible with embodied, enactive and extended conceptions of cognition~\citep{clark2017knit}, it is possible to construct such a dependency structure for (F, A)-systems. Such a construction is potentially compatible with other recent models that have been proposed to understand agency~\citep{biehl2022interpreting, kiverstein2022problem, egbert2023behaviour}. These detailed explorations are beyond the scope of this paper. 

Here, it suffices to note that (F, A)-systems inherently embody a form of anticipation. By integrating mechanisms for sensing and response through freestanding formal causes, and embedding them within asynchronous, multi-scale temporal structures, the system develops the capacity to interpret environmental inputs before they fully unfold. This anticipatory structure resonates with Rosen’s conception of anticipatory systems as those that contain internal models capable of projecting future states and acting upon those projections in the present~\citep{rosen1978anticipatory}. 

In Rosen’s formal terms, such systems exhibit a predictive model whose outputs serve not merely as representations, but as causal drivers of current behavior, apparently violating classical notions of causality, yet foundational for biological adaptivity~\citep{rosen1978anticipatory}. Our temporally extended (F, A)-systems instantiate this idea constructively: rather than relying on retrospective feedback, the system deploys present environmental cues to actively constrain future states through model-mediated inference. This marks a shift from mere reaction to primitive interpretation. Since we are showing that a genuine process of interpretation takes place within simple life forms, such as a cell, our extension of (F, A)-systems reconciles Barbieri's code biology~\citep{barbieri2019code} with Pattee's physical biosemiotics~\citep{pattee2001physics}. The following section is devoted to the study of weak versions of the relational model developed thus far to explore the origin and evolution of semantic closure.

\section{Deriving Weaker Models for Primitive Semantic Closure}
From our previous analysis, we know that diagram~(\ref{d:hofmeyr_full_2}) accounts for semantic closure. Having found a minimal model with this feature, we will now derive weaker versions of diagram~(\ref{d:hofmeyr_full_2}) to 
glimpse the evolutionary emergence of semantic closure.

\begin{equation}
\label{d:hofmeyr_full_2}
\begin{tikzcd}[column sep=0 cm]
& X \ar[r,thick] & Y \ar[r,thick] & Z \\
&& I \times \{f_1\}+f_3+f_4+f_2 \ar[lld,dashed,thick,start anchor={[xshift=-4ex]}]
\ar[rr,dashed,thick] \ar[ul,dashed,thick,start anchor={[xshift=3ex]}]
\ar[u,dashed,thick,start anchor={[xshift=4ex]}] & & D \ar[d,thick]   \\
A \ar[rr,thick] & & B_1+B_2+B_3+B_4 \ar[u,thick] & & \Phi \ar[ll,thick]
\end{tikzcd}
\end{equation}

\citet{pattee2012does} discusses how hierarchical levels of control in biology emerge through evolution. In early life, simple functional switches (e.g. enzyme activation or inhibition) would allow a system to respond to internal changes, manage its own metabolic processes, and gradually evolve toward more complex forms of environmental interaction, where sensing and adaptation to external stimuli would be added later. Thus, our first step to proceed down our logical chain of assumptions and produce a weak version of the diagram (\ref{d:hofmeyr_full_2}), is to eliminate $f_3+f_4$ and $B_3+B_4$, obtaining the diagram (\ref{d:hofmeyr_full}) shown below.

\begin{equation}
\label{d:hofmeyr_full}
\begin{tikzcd}
  & I^*\times\{f_1\}+f_2 \arrow[dl, dashed, thick] \arrow[r, dashed, thick]  & D \arrow[d, thick] \\
A \arrow[r, thick] & \arrow[u, thick] B_1+B_2 & \arrow[l, dashed, thick] \Phi_f  
\end{tikzcd}
\end{equation}

In diagram (\ref{d:hofmeyr_full}), $I^* = \{I_1, I_2\}$. Although digital information processing is a fundamental requirement for life~\citep{neumann1966theory,waters2012neumann, marletto2015constructor}, according to~\citet{pattee2012does} digital messengers emerge as a product of a synergistic process, akin to how formal rules of syntax and dictionaries evolved from the functional use of primitive symbols in a complex environment. Along these lines, digitization would have arisen as a natural outcome of the processes in reaction networks that had once been primarily analog. This implies that $I^* = \{I_1, I_2\}$ in diagram~(\ref{d:hofmeyr_full}) could not have existed at the beginning. 

Since early chemical systems relied only on geophysical and geochemical constraints, with no external formal causes or instructions~\citep{pattee2012does}, there are two possibilities to resolve infinite regress~\citep{hofmeyr2018causation, pattee2021symbol}. The first is that $f_1$ is an \emph{informed} efficient cause (a semantic interpreter whose specificity is embedded in it), as $f_2$ already is. The second is that $A$, although primarily a material cause, assumes the role of an \emph{informed} material cause (a symbolic substrate whose syntax is materially inscribed), similar to $B_1$ and $B_2$~\citep{hofmeyr2021biochemically}. The second option represents the most plausible evolutionary pathway for implementing a weaker version of our (F, A)-system~\citep{pattee2021symbol}. This leads us to the successor diagram below:

\begin{equation}
\label{d:hofmeyr_succesor}
\begin{tikzcd}
& f_1+f_2 \arrow[dl, color=black, dashed, thick] \arrow[r, color=black, dashed, thick]  & D \arrow[d, color=black, thick] \\
A_1^* + A_2^* \arrow[r, color=black, thick] & \arrow[u, color=black, thick] B_1+ B_2 & \arrow[l, color=black, dashed, thick] \Phi_f   
\end{tikzcd}
\end{equation}

\textcolor{blue}{In diagram~(\ref{d:hofmeyr_succesor}), $A_1^*$ and $A_2^*$ represent informed material causes, $f_1$ maps the disjoint set $A_1^* + A_2^*$ to the disjoint set $B1 + B2$, and $f_2$ maps $D$ to $\Phi_f$.} This evolutionary pathway relies on intrinsic constraints and properties rather than complex external mechanisms. Prior to this stage, however, $f_2$ must be replaced by a non-informed efficient cause ($f_2^*$), and $D$ must transition into an informed material cause ($D^*$):

\begin{equation}
\label{d:hofmeyr_ss}
\begin{tikzcd}
& f_1+f_2^* \arrow[dl, color=black, dashed, thick] \arrow[r, color=black, dashed, thick]  & D^* \arrow[d, color=black, thick] \\
A_1^* + A_2^* \arrow[r, color=black, thick] & \arrow[u, color=black, thick] B_1+ B_2 & \arrow[l, color=black, dashed, thick] \Phi_f   
\end{tikzcd}
\end{equation}

In diagram~(\ref{d:hofmeyr_ss}), $D^*$ acts as an informed material cause that is prone to being transformed by $f_2^*$ into $\Phi_f$. This architecture retains the autonomy and hierarchical closure of the system. If $\Phi_f = \Phi_{f_1} + \Phi_{f_2}$ and $B_2$ is provided by the environment (and therefore does not have to be made from $A_2^*$) then diagram~(\ref{d:hofmeyr_ss}) can be split into two components~\citep{hofmeyr2021biochemically}. The first component is:

\begin{equation}
\label{d:component1}
\begin{tikzcd}
& f_1 \arrow[dl, color=black, dashed, thick] &  \\
A_1^* \arrow[r, color=black, thick] & \arrow[u, color=black, thick] B_1 & \arrow[l, color=black, dashed, thick] \Phi_{f_1}
\end{tikzcd}
\end{equation}

Although this component is not closed to efficient causation, given that it has the form of a non-replicating (M, R)-system, the states and inputs cannot be separated, so it cannot be instantiated as a finite-state automaton \citep{rosen1964abstract}. We can represent diagram~(\ref{d:component1}) as the following hierarchical cycle: 

\begin{equation}
\label{d:c1-hc}
\begin{tikzcd}
& \Phi_{f_1} \arrow[dl, color=black, dashed, thick]  \\
B_1 \arrow[r, color=black, thick] & f_1 \arrow[dl, color=black, dashed, thick]\\   
A_1^* \arrow[r, color=black, thick] & B_1
\end{tikzcd}
\end{equation}

As~\citet{palmer2016rosen} explained, such a hierarchical cycle can be represented by an automaton with memory. We can then further split diagram~(\ref{d:c1-hc}) into

\begin{equation}
\label{d:c1-1}
\begin{tikzcd}
& \Phi_{f_1} \arrow[dl, color=black, dashed, thick]  \\
B_1 \arrow[r, color=black, thick] & f_1 
\end{tikzcd}
\end{equation}

and

\begin{equation}
\label{d:c1-2}
\begin{tikzcd}
& f_1 \arrow[dl, color=black, dashed, thick]  \\
A_1^* \arrow[r, color=black, thick] & B_1 
\end{tikzcd}
\end{equation}

Both diagrams~(\ref{d:c1-1}) and~(\ref{d:c1-2}) can be represented using a finite state automaton~\citep{rosen1964abstract, palmer2016rosen}. The second component of diagram~(\ref{d:hofmeyr_ss}) is:

\begin{equation}
\label{d:component2}
\begin{tikzcd}
f_2^* \arrow[r, color=black, dashed, thick]  & D^* \arrow[d, color=black, thick] \\
\arrow[u, color=black, thick] B_2 & \arrow[l, color=black, dashed, thick] \Phi_{f_2}   
\end{tikzcd}
\end{equation}

Unlike the first, this component is still closed to efficient causation and thus semantically closed. A rather pertinent observation is that there is a clear parallel between diagram (\ref{d:component2}) and the \emph{autogenic virus} (“autogen” for short) proposed by \citet{deacon2021molecules}. Autogen arises from the interaction of two self-organizing processes: \emph{reciprocal catalysis} and \emph{self-assembly}, the former producing locally asymmetrically high concentrations of a small number of molecular species and the latter requiring persistently high local concentrations of a single species of component molecules. Autogen's molecular logic is summarized by diagram (\ref{d:autogen}).

\begin{equation}
\label{d:autogen}
\begin{tikzcd}
f + g \arrow[r, color=black, dashed, thick]  & a \arrow[d, color=black, thick] \\
\arrow[u, color=black, thick] d & \arrow[l, color=black, dashed, thick] n+c  
\end{tikzcd}
\end{equation}

In diagram (\ref{d:autogen}) $a$ and $d$ are material substrates provided by the environment. In addition, $f$ and $c$ are molecular catalysts, while $n$ and $g$ are by-products of the chemical interaction, where $g$ promotes self-assembly~\citep{deacon2021molecules}. Since in diagram~(\ref{d:component2}) $f_2^*$ represents the maintenance of the intracellular milieu $\Phi_{f_2}$ that makes self-assembly possible~\citep{hofmeyr2021biochemically}, we can equate $f+g\triangleq \Phi_{f_2}$. Thus, $n+c \triangleq f_2^*$, $a \triangleq B_2$, and $d \triangleq D^*$. In this way, \citet{deacon2021molecules}'s autogen is a particular case of the model presented here. Returning to diagram (\ref{d:component2}), before achieving closure to efficient causation, the chemical reaction network must have been open. This can be represented as:

\begin{equation}
\label{d:no_closed}
\begin{tikzcd}
& \Phi_{f_2} \arrow[dl, color=black, dashed, thick]  \\
B_2 \arrow[r, color=black, thick] & f_2^* \arrow[dl, color=black, dashed, thick]\\   
D^* \arrow[r, color=black, thick] & E
\end{tikzcd}
\end{equation}

or

\begin{equation}
\label{d:no_closed_2}
\begin{tikzcd}
& f_2^* \arrow[dl, color=black, dashed, thick]  \\
D^* \arrow[r, color=black, thick] & \Phi_{f_2} \arrow[dl, color=black, dashed, thick]\\   
B_2 \arrow[r, color=black, thick] & E
\end{tikzcd}
\end{equation}

In diagrams~(\ref{d:no_closed}) and (\ref{d:no_closed_2}), $E$ simply represents another chemical product. In both cases, this system is no longer closed to efficient causation, allowing its hierarchical entailments to be modeled using a stream X-machine~\citep{palmer2016rosen}. When the functional entailments are further dismantled, simpler components acquired the form~(\ref{d:c1-1})-(\ref{d:c1-2}) and can be represented by finite state automata~\citep{rosen1964abstract}. From the above we can glimpse a clear transition across the computational capabilities from simple chemical reactions to quasi-autonomous organisms, such as our initial extended (F, A)-system. Based on our theoretical observations, in the next section we propose an evolutionary narrative in computational enactivist terms~\citep{korbak2021computational} for the origin of semantic closure. As we shall see, such an account is consistent with the experimental results obtained by~\citet{duenas2019chemistry}.

\section{Towards a Computational Enactivist Narrative for the Origin and Early Evolution of Semantic Closure}
Following the line of thought proposed by~\citet{pattee2012does}, the primordial ecosystem can be conceptualized as the first information-containing architecture on Earth, where primitive geochemical cycles gave rise to a well-defined syntax (based on chemistry) spontaneously and without requiring genetic instructions or initial metabolic control. Given a set of primordial geophysical and geochemical constraints---a primeval \textit{ecosystem language} \citep{pattee2012does}---we can envisage the emergence of simple bimolecular reactions, such as those represented in diagrams~(\ref{d:c1-1}) and~(\ref{d:c1-2}), which can be realized as finite-state automata. These simple chemical reactions are capable of recognizing regular languages, a task that does not require counting or memory~\citep{duenas2019chemistry}. 

In this primitive world, molecules proficient in catalyzing key environmental reactions and processes became abundant over time~\citep{kuppers1990information}. Later, the diversity of such molecules allowed interactions and, in some cases, bonding, leading to systems of multiple interconnected reactions mediated by common intermediate chemical species. In this panorama, some intermediates functioned both as the product of one reaction and as the reactant for another, forming networks of increasingly complex chemical reactions. These ideas align with the emergence of composomes \citep{segre2000composing, hunding2006compositional} and their maintenance \citep{markovitch2012excess}.

Such complex reactions can be represented through hierarchical functional entailments that are not closed to efficient causation (see diagrams~(\ref{d:c1-hc}), (\ref{d:no_closed}), and (\ref{d:no_closed_2})) and can be realized using stream X-machines, as suggested by~\citet{palmer2016rosen}. This primitive form of memory based on the reuse of catalytic elements enables the recognition of more complex sequences of information~\citep{hopcroft2001introduction} and enhances the evolutionary robustness of such information under changing environmental conditions~\citep{walker2013algorithmic}. Despite their simplicity, these chemical reactions could encode correlations in environmental variables~\citep{bartlett2022provenance} and potentially exhibit a basic form of anticipation~\citep{poole2017chemical}.

This primeval landscape helps outline the evolutionary trialectic proposed by~\citet{jaeger2024naturalizing}, where the environment plays a crucial role in shaping the catalytic abilities of chemical species via random mutations induced by geochemical and geophysical processes. This activity fosters molecular Darwinism~\citep{kuppers1990information}, where the regulation of terrestrial cycles and adaptation to abrupt changes become, respectively, the goals and actions of this transjective trialectic. Furthermore, this competition among chemical components also fosters cooperative homeostatic loops that improve predictive capabilities and establish functional relationships~\citep{levin2019computational, heins2024collective}.

Chemical reactions can become increasingly complex either through the accumulation of catalytic mutations or by coupling chemical entities of varying complexity. Either pathway allows the emergence of simple autocatalytic processes represented by hierarchical cycles semantically closed (see diagram~(\ref{d:component2})), equivalent to \citet{deacon2021molecules}'s autogen. The impredicative nature of any system closed to efficient causation can be interpreted as nested symbolic loops of computation---i.e., self-reference---that are transmuted into non-symbolic processes of self-replication~\citep{sayama2008construction}. In this light, self-replication constitutes the threshold for open-ended evolution~\citep{hernandez2018undecidability}, wherein the evolutionary process itself becomes increasingly unconstrained and generative~\citep{pattee2019evolved}. Evolution, therefore, emerges as a fundamentally syntax-pragmatic phenomenon, characterized by the unpredictability and emergence of future states. An insight consistent with the trialectic framework articulated by \citet{jaeger2024naturalizing}.

A distinction must be drawn between self-replication and self-reproduction. Self-reproduction requires a copier $C$ acting on raw materials~\citep{szathmary1995classification}, whereas in self-replication, the copier $C$ is null, being implicit in the laws of physics~\citep{marletto2015constructor}. Examples of self-replication include crystals, short RNA strands, and autocatalytic cycles. In resource-limited environments, chemical networks compete, self-replicate, and mutate through minor variations in composition, ensuring that each type of replicator produces at least one viable offspring, on average, per lifetime. This constitutes a primitive form of evolution~\citep{marletto2015constructor}, characterized by natural selection with limited inheritance due to low robustness and accuracy in replication. 

Another important consideration is that, even at this primordial stage, geophysical and chemical processes are naturally cyclical, ensuring a constant energized, non-equilibrium, dynamic exchange between environment and self-replicators. Under these conditions, dynamic kinetic stability (DKS), a kinetic form of stability associated with entities capable of self-replication, emerges~\citep{pascal2015stability}. DKS guides self-replication toward greater robustness~\citep{pascal2023toward}, enabling increasingly complex dynamics to arise as self-replicators proliferate~\citep{alakuijala2024computational}, preserving encoded environmental information and preventing its deterioration.

As we pointed out before, the syntactic, semantic, and pragmatic dimensions of information are inseparable. Although the presence of semantic (or pragmatic) information in such a primordial state may seem improbable, meaning operates at multiple scales, being syntax-pragmatic interactions its only requirement for existence~\citep{fields2020living}. According to~\citet{kuppers1990information}, the protosemantics of biological information is defined by the carrier’s capacity to reproduce rapidly while maintaining high accuracy and stability. This aligns with DKS and likely facilitated the development of sophisticated semantic closure~\citep{pattee2012evolving}.

Over time, the process of infotaxis likely also facilitated compartmentalization, enabling chemical networks to associate mutually, enhancing their functionality and robustness. This progression can be represented by diagrams like~(\ref{d:hofmeyr_ss}), where formal causes are inherent within material causes~\citep{hofmeyr2018causation}. As these networks became more complex, they began to evolve formal causes inherent to efficient causes, such as $f_2$ in diagram~(\ref{d:hofmeyr_succesor}). However, purely analog systems exhibited limited robustness and evolutionary potential compared to analog systems supplemented by digital information control~\citep{walker2013algorithmic}. 

Digitization emerged as a natural outcome of the evolution of analog reaction networks, eventually resembling systems like diagram~(\ref{d:hofmeyr_full}), where $I^*$ acts as a freestanding formal cause. This transition allowed chemical networks to communicate more efficiently among themselves and with their progeny, being capable of developing sophisticated mechanisms to sense and response to the environment, as shown in diagram (\ref{d:hofmeyr_full_2}). However, the emergence of truly open-ended evolution required more than enhanced signaling or compartmentalized inheritance. It demanded the appearance of systems capable of constructing new functional configurations from internally stored, symbol-like instructions, thus enabling not merely replication, but the continuous rewriting of what it means to persist. Such systems exhibit a form of universality grounded in discrete, modular, and syntactically governed structures; codes that are not just molecularly embodied but materially inert, enabling symbolic reuse across vastly different contexts~\citep{rocha2005material}. 

As these symbolic capacities expanded, the biosphere itself began to resemble a distributed, asynchronous computation over possible adaptations: not in the sense of simulating all paths, but in pruning the unviable ones through irreversible commitments like extinction~\citep{hernandez2018undecidability}. In this setting, survival is not merely the preservation of form, but the preservation of a semantic loop. A closure between symbolic description, constructed function, and environmental viability. What emerges is not just complexity, but a form of undecidable elaboration: a dynamical process in which the criteria for persistence are themselves continuously evolving, grounded in the dual logic of physical constraint and symbolic freedom.

Thus, we observe a continuum of increasing complexity, beginning with simple chemical reactions modeled as finite state automata (see diagram~(\ref{d:component1})) and culminating in organisms with cellular-like architectures, such as the extended (F, A)-system depicted in diagram~(\ref{d:hofmeyr_full_2}), which exhibit nested loops of computation. In this way, our model explains the experimental observations made by~\citet{duenas2019chemistry}. Digitization would have been a gradual process, taking substantial time to unfold. Ultimately, digital life forms may have become dominant, as they are intrinsically more robust and better equipped to survive in the long term, representing the lasting products of life’s evolutionary processes~\citep{walker2013algorithmic}. 

\section{Discussion}
Through our theoretical model, we have outlined a sequential and hierarchical pathway for the origin and early evolution of semantic closure in nature. As noted by~\citet{jaeger2024naturalizing}, relevance realization is a process that cannot be fully captured algorithmically. Our narrative complements this view by situating autopoiesis and adaptation within a relational model that is closed semantically while remaining open syntactically. A simple temporal unfolding of this model equips even the simplest life forms with anticipatory mechanisms, validating autopoiesis, adaptation, and anticipation based on active inference as syntax-pragmatic mutualistic processes~\citep{pattee2007necessity}. 

While~\citet{jaeger2024naturalizing} highlight the apparent non-computability of relevance realization, their conception of computation centers around von Neumann architecture and Turing machines, ignoring computing architectures capable of implementing measurement and control in their operation, which could reveal learning processes mirroring those observed in living systems. We think this narrative is more meaningful through the lens of computational enactivism~\citep{korbak2021computational}. This perspective anchors the active inference mechanisms of organisms in the temporal parameterization of our relational model, taking into account both syntactic (computational) and pragmatic (enactive) aspects of life.

Our model successfully captures numerous properties characteristic of living systems, with the notable exception of controlled growth~\citep{cornish2020contrasting, cornish2022essence}. However, a recent study proposes a model inspired by Rosen's original view of (M, R)-systems, capturing life as bipartite directed multigraphs~\citep{marquez2025nature}. This approach allows modeling controlled growth. As future work, one idea might be to relate their perspective with ours. Crucially, our model is flexible and independent of specific conditions, being agnostic to the substrate, period, or location in which life might have emerged~\citep{abbot2011steppenwolf, loeb2014habitable,pearce2018constraining,ballesteros2019diving}. The only fundamental environmental prerequisite in our framework is the presence of a primordial language that enables the processes underlying life to take root. This perspective aligns with the idea that symbol grounding is a necessary precursor to interpretation, as emphasized by~\citet{pattee2021symbol}.

As~\citet{cardenas2018rosennean} demonstrated, the absence of strict hierarchies in organisms enables closure to efficient causation in relational models, a principle upheld in our extended (F, A)-system. This makes circular causality a cornerstone of our narrative, which is fully compatible with the principle of biological relativity~\citep{noble2012theory, noble2019biological}. Biological relativity posits that no single level of causation is privileged to describe life, ruling out phenomena such as downward causation. Instead, this aligns with the concept of causal spreading, which suggests that any component within a biological system can influence any other, depending on context~\citep{ball2023distinguishes}.

While our narrative illustrates how primitive mechanisms of information control in life might have evolved, it does not claim to pinpoint the precise moment life began. Depending on the definition we use~\citep{benner2010defining}, life could be said to arise at various points in our framework. From a gradualist perspective, this ambiguity is not problematic~\citep{kuppers1990information, gershenson2012world}. However, it is clear that after the emergence of structures closed to efficient causation (and, as a consequence, semantically closed), many of the defining properties of life emerged. This marks what we might call a \emph{self-referential takeover}, as chemical systems beyond this point acquire nested loops of computation. Our analysis demonstrates that self-referentiality is not only a necessary condition for life but also essential even for the simplest autocatalytic systems.

\section{Conclusion}
In this paper, we extended the (F, A)-systems proposed by~\citet{hofmeyr2021biochemically} through a temporal parametrization. This extension captures not only fundamental properties of life, such as thermodynamic openness, catalytic specificity, catalytic and structural closure, information handling, reproduction, and evolvability, but also anticipation. Using this relational biological model, we systematically derived weaker versions that reveal a hierarchy of processes by which chemical systems acquired increasingly advanced computational (syntactic) and enactive (pragmatic) capabilities, consistent with the experimental findings of~\citet{duenas2019chemistry}. 

Semantic closure emerges as a key factor for robustness, self-replication, and thus the open-ended evolutionary dynamics observed in living systems. Furthermore, by attributing a self-referential nature to life, our computational enactivist framework aligns with~\citet{jaeger2024naturalizing}'s resolution of the problem of relevance. While life (and its consequences, such as agency and cognition) cannot be represented by von Neumann architectures or Turing machines, our work encourages the exploration of architectures that explicitly implement measurement, control, and stochasticity to realize life.

Some problems remain open for future exploration. First, the formulation of a Markov blanket for the Asynchronous Dynamic Bayesian Network associated with our temporal parametrization could provide critical insights into relationships among emerging theoretical frameworks aimed at constructing a general theory of agency~\citep{egbert2023behaviour}, providing a concrete model for the sensorimotor autonomy described by~\citet{kiverstein2022problem}. Second, similarly to what we did with our extended (F, A)-systems, we can derive the temporal parametrizations and joint distributions for weak versions of our model, glimpsing the emergence of internal models that guide adaptive action not just in (M, R)-systems, but across the entire zoo of systems closed to efficient causation discussed by~\citet{hofmeyr2021biochemically}. Third, once a taxonomy of semantically closed systems has been provided, we could examine the consequences of having two or more relational entities occupying the same environment, forming an ecology where multiple cellular-like systems can be coupled to build multicellular organisms. 

Our work represents a step towards the unification of three intellectual traditions: relational biology, physical biosemiotics, and ecological psychology; each of which fundamentally challenges mainstream reductionist and mechanistic views of life and cognition. Intriguingly, all three paradigms originated within the intellectual landscape of the Northeastern United States, reflecting a shared historical and geographical milieu that fostered a relational, systemic, and meaning-centered conception of living systems. By situating our extended (F, A)-systems at the intersection of these traditions, we have articulated a cohesive theoretical synthesis, tentatively termed “The Northeastern Synthesis,” highlighting the profound relevance of relational autonomy, semiotic processes, and affordance realization in biological information handling. In line with~\citet{stewart1995cognition}'s constructivist perspective, we propose that our narrative has the potential to explain a wide range of phenomena, from the origin of life to high-level cognition.

\section{Author Contributions}
A.J.L.-D.: formal analysis, visualization, conceptualization, investigation, methodology, validation; C.G.: conceptualization, supervision, funding acquisition, validation.

\section{Acknowledgements}
The authors acknowledge Jan-Hendrik S. Hofmeyr, Hiroki Sayama, Juan Pérez-Mercader, Kate Nave, Luis M. Rocha, Cliff Joslyn, Dennis P. Waters, and Pedro Márquez-Zacarías for stimulating conversations and their valuable comments on earlier versions of this manuscript, as well as anonymous reviewers for constructive comments when submitting this work.

\section{Data Accessibility}
This article has no additional data.

\section{Funding Statement}
C.G. acknowledges support from the SSIE School.

\footnotesize
\bibliographystyle{apalike}
\bibliography{example} 

\printglossaries

\end{document}